\documentclass[]{jsedi_pub}

\title{The mantle-inner core gravitational mode of oscillation in a strong magnetic field regime}
\shorttitle{The MICG mode in a strong magnetic field regime} 

\author[1]{Mathieu Dumberry
	\orcid{0000-0001-5677-1582}
	\thanks{Corresponding author: 
    \href{mailto:dumberry@ualberta.ca}{dumberry@ualberta.ca}}
}

\affil[1]{Department of Physics, University of Alberta, Edmonton AB, Canada}

\begin{document}


\publicationonly{
\dois{10.46298/jsedi.15735}
\handedname{Alexandre Fournier}
\receiveddate{May 23, 2025}
\accepteddate{September 23, 2025}
\publisheddate{September 26, 2025}
\theyear{2025}
\thevolume{1}
\thepaper{1}  
}

\makesedititle{
  \begin{summary}{Abstract}
The mantle-inner core gravitational (MICG) mode is the free mode of axial oscillation between the mantle and inner core, sustained by the gravitational torque between their degree 2 order 2 density structures. As part of this mode, the tangent cylinder (TC) is entrained to move jointly with the inner core, and the oscillations of the TC launch Alfv\'en waves propagating in the region outside the TC.  Here, we investigate how the MICG is altered by the strength of the internal magnetic field in the core which controls the travelling speed of Alfv\'en waves.  We show that the MICG mode remains a distinct normal mode of oscillation of the core-mantle system only when Alfv\'en waves are attenuated before they traverse the width of the fluid core.  For an internal magnetic field strength of a few mT, as we expect in Earth's core, Alfv\'en waves can readily traverse the width of the core, and the MICG mode is absorbed into the spectrum of torsional oscillation (TO) modes.  The MICG period retains a dynamical influence, acting as a point of resonance for TO modes, and marking the transition from a TO mode that is weakly impacted by gravitational coupling to one in which the motion of the TC (including the inner core) is strongly restricted. Our results imply that the observed 6-year periodic signal in the length of day cannot be interpreted as the signature of the MICG mode and must instead be caused by TO modes, or more generally, by the propagation of Alfv\'en waves.   
   \vspace{1.3cm} 
  \end{summary}
  }
 \begin{summary}{Non-technical summary}
The Earth's mantle is not uniform but contains continental size density anomalies.  The topography of the solid inner core (located at the centre of our planet) is aligned on average with these mantle density anomalies. The axial gravitational restoring torque between these mantle and inner core structures -- when they are rotated out of longitudinal alignment -- sustains a natural mode of oscillation termed the mantle-inner core gravitational (MICG) mode.  Previous studies have suggested that the MICG mode may explain a periodic signal of 6 years observed in the changes in the length of day (LOD).  Here, we show that when the magnetic field inside the fluid core is high (of the order of a few mT), the nature of the MICG changes; it is no longer a distinct, independent mode of oscillation of the core-mantle system.  Instead, the MICG mode merges into the set of modes of torsional oscillations (TO), which consist in the longitudinal oscillations of nested cylindrical surfaces within the fluid core maintained by the magnetic tension between them (oscillations that are also known as Alfv\'en waves).  The inner core can still oscillate with respect to the mantle, but it is due to its entrainment as part of the TO modes. These results have important implications for the interpretation of the 6 year LOD signal, which cannot be caused by the MICG mode, but must instead be caused by the TO modes, or more generally, by the propagation of Alfv\'en waves in the fluid core.   
 \end{summary}

\section{Introduction}

The mantle-inner core gravitational (MICG) mode of axial oscillation owes its existence to aspherical mass anomalies within the mantle, including those caused by the topography at surfaces of density contrast \citep{buffett96a}.  The largest of these occur at spherical harmonic degree 2 order 2  \citep[e.g.][]{defraigne96,simmons07}, warping equipotential surfaces everywhere inside the Earth (including the surface geoid) into ellipsoid shapes elongated in the equatorial plane.  Within the core, surfaces of constant density coincide with these elliptically shaped equipotential surfaces in order to maintain hydrostatic equilibrium.  Provided the inner core deforms viscously on a timescale shorter than the typical changes in the gravitational potential imposed by the mantle, the inner core boundary (ICB) should also match an equipotential surface.  Its shape should then be elliptical in the plane of the equator (i.e. equatorially elliptical) and longitudinally aligned with the leading order mantle mass anomalies \citep{buffett96a}.  An axial rotation of the inner core out of this longitudinal alignment, driven for example by magnetic stresses at the ICB \citep[e.g.][]{buffett00,aubert11}, leads to a gravitational torque acting to restore the equilibrium state of minimum gravitational potential energy \citep{buffett96a}.  This gravitational torque can thus sustain a natural mode of axial oscillation between the inner core and the mantle, referred to as the MICG mode \citep{mound06}. 

The period of the MICG mode depends on the gravitational strength factor, $\overline{\Gamma}$, capturing the amplitude of the gravitational torque for a given small longitudinal misalignment between the inner core and mantle \citep{buffett96b}.  In turn, $\overline{\Gamma}$ depends on the amplitude of the degree 2 order 2 geoid at the core-mantle boundary (CMB), which itself depends on estimates of the mantle density anomalies, including that caused by the topography of the CMB \citep{buffett96a,davies14,chao17,shih21}. The period of the MICG mode also depends on whether the region of the fluid core inside the axial cylinder that encloses the inner core -- the so-called tangent cylinder (TC) -- is entrained into co-rotation with the inner core, which is expected given the strong electromagnetic coupling at the ICB \citep{buffett96b,mound03}.  

Estimates of the period of the MICG mode can be derived by computing predictions of $\overline{\Gamma}$ on the basis of viscous mantle flow models.  The latter are built from density anomalies within the mantle inferred from seismic tomography and the subsequent viscous deformation of surfaces of density discontinuities that they cause \citep[e.g.][]{hager85,forte94,defraigne96}.  Based on the mantle flow models available at the time, \citet{buffett96a} estimated a MICG period of approximately 2 to 3 years. The more recent survey of  \citet{davies14} suggests a longer period in the range 7 to 18 years.  The range of uncertainty reflects our limited knowledge of both the amplitude of the internal mantle density anomalies \citep[primarily those of degree 2, order 2 associated with Large Low Seismic Velocity provinces in the lower mantle, e.g.][]{garnero16,mcnamara19} and the viscosity of the lowermost mantle.

An alternate approach is to seek the signature of the MICG mode in the observed changes in the mantle rotation rate measured as changes in the length of day (LOD).  Indeed, an observed quasi-periodic LOD signal with an amplitude of $\sim 0.13$ ms and a period of $\sim 5.9$ years \citep{abarcadelrio00,holme13,chao14} has been linked with the MICG mode \citep{mound06,chao17,ding18,duan18,shih21}.  The magnitude of $\overline{\Gamma}$ required to match this period provides then a direct measure of the equatorial ellipticities of the geoid at the CMB and of the ICB topography \citep{davies14,chao17,shih21}.
  
This simplified picture of the MICG mode neglects the dynamical influence of the fluid core in the region outside the tangent cylinder (ROTC) or assumes that it plays a passive role.  At periods relevant to the MICG mode, axisymmetric azimuthal (zonal) flows are expected to be invariant in the direction of the rotation axis and take the form of rigidly rotating nested cylindrical surfaces \citep{jault08}.  A differential motion between adjacent cylinders shears the cylindrically radial ($B_s$) magnetic field lines that threads their surfaces.  This creates an azimuthal restoring force between cylinders and allows for the propagation of Alfv\'en waves in the cylindrically radial direction \citep{braginsky70}. 

The normal modes of axial oscillations between the cylinders formed by the superposition of these Alfv\'en waves are referred to as torsional oscillations \citep[TO,][]{braginsky70} and their periods depend the strength of $B_s$ within the core. These TO modes include the motion of the TC, so they are necessarily coupled to the MICG mode.  The prevalent view at the time the earliest studies on the MICG mode were published \citep[e.g.][]{buffett96a,buffett96b,mound03,mound06} was that the gravest mode of TO had a period of 60-80 years, implying a $B_s$-field strength of 0.2-0.3 mT within the core \citep{zatman97}.  Under this scenario, the MICG period is distinctly shorter than that of the fundamental TO mode.  Consequently, the MICG mode provides a resonant band of amplification to higher TO harmonics that have periods in the vicinity of the MICG period \citep{mound03}. The oscillating motion of the inner core, entrained by the TO modes, exerts a gravitational torque on the mantle leading to distinct peaks in the LOD at periods close to the MICG mode.  Even though the ROTC is dynamically involved, it carries little angular momentum and the primary angular momentum balance for these amplified TO modes is between the mantle and the whole of the TC, just as is the case for a pure MICG mode.  The 5.9 yr oscillation in the LOD may still be tied to MICG period and provides a constraint on $\overline{\Gamma}$ \citep{mound06}.

The general consensus on the magnetic field strength within the core has since shifted.  Numerical models of the dynamo suggest that the field deep inside the core is likely stronger by a factor 10 than at the CMB, suggesting an internal field strength of 3-4 mT \citep[e.g.][]{christensen06,aubert09}.  With such a large field strength, Alfv\'en waves propagate through the entire width of the core in only a few years and the periods of the gravest TO modes are expected to be in the subdecadal range. Indeed, \citet{gillet10} showed that the angular momentum carried by propagating Alfv\'en waves matches that required to explain the observed periodic $5.9$-yr LOD signal. This indicates that the mantle exchanges its angular momentum with the whole of the core rather than with the TC alone, in contrast with an idealized MICG mode. Furthermore, in this scenario, the 5.9 yr period is instead tied to the period of the fundamental mode of TO, not the strength of the gravitational torque, even though the gravitational torque may still provide the means by which angular momentum is transferred to the mantle.  

In a strong $B_s$-field regime, Alfv\'en waves travel through the width of the core in a shorter time than the MICG period.  Here, our goal is to investigate how the MICG mode is altered by such fast propagating Alfv\'en waves.  We shall show that when Alfv\'en waves can propagate through the ROTC before they get attenuated, the MICG mode is no longer an independent mode of the coupled core-mantle system.  Instead, it gets absorbed into the spectrum of TO modes, although the latter are affected by gravitational coupling. 

\section{Theory}

\subsection{Angular momentum balance}

The model of internal coupling between the mantle, fluid core and inner core that we use follows broadly that presented in several studies, including in \citet{buffett96b,buffett98,mound03,mound05b,dumberry08c,dumberry10b}.  We assume a spherical model of Earth's interior comprised of an inner core (radius $r_i$), a fluid core (outer radius $r_f$) and mantle.  We adopt a simplified model of flow in the fluid core restricted to its azimuthal component $v_\phi$ and assumed to be axisymmetric and invariant along the direction of the rotation axis.  Accordingly, we use cylindrical coordinates ($s$, $\phi$, $z$), and we divide the fluid core in a set of $N_s$ discrete cylindrical surfaces with radius $s$ that intersect the CMB at axial position $z_f=\sqrt{r_f^2 - s^2}$ in the Northern hemisphere.  Cylinders inside the TC ($s<r_i$) intersect the ICB at $z_i=\sqrt{r_i^2 - s^2}$; we set $z_i=0$ for cylinders outside the TC. The azimuthal velocity of each cylinder is set equal to $v_\phi = s \Omega_f$, where  $\Omega_f \equiv \Omega_f(s)$ is the angular velocity.  

In the absence of an external torque, the axial angular momentum equations of the mantle, inner core and fluid core are given, respectively, by
\begin{subequations}
\begin{align}
 C_m \frac{d}{dt} \Omega_m  & =  \overline{\Gamma} \alpha + \Gamma_{cmb}  \, , \label{eq:tqm} \\
 C_i \frac{d}{dt} \Omega_i & = - \overline{\Gamma} \alpha + \Gamma_{icb}   \, ,\label{eq:tqi} \\
 \int_0^{r_f} c_f \,  \frac{d \Omega_f}{dt} \, ds & = -\Gamma_{cmb}  -\Gamma_{icb} \, ,\label{eq:tqf} 
\end{align}
\end{subequations}
where, $C_m$, $C_i$ are the axial moments of inertia of the mantle and inner core, $\Omega_m$, $\Omega_i$ their rotation rates, and
\begin{equation}
c_f = 4 \pi \rho s^3 (z_f - z_i) \, ,
\end{equation}
is the axial moment of inertia density of cylinders within the fluid core, where $\rho$ is the density (assumed uniform).  $\Gamma_{cmb}$ and $\Gamma_{icb}$ are the torques from all surface forces acting on the mantle at the CMB and on the inner core at the ICB, respectively.  $\overline{\Gamma}$ is the strength of the gravitational torque between the mantle and inner core and $\alpha$ is the (assumed small) longitudinal angle of misalignment between the degree 2 order 2 mantle density field and the long equatorial axis of the ICB topography.  $\alpha$ depends on the differential rotation between the inner core and the mantle and on the viscous relaxation time ${\tau}$ for the ICB topography to return to its equilibrium alignment with the imposed gravitational potential from the mantle; its evolution is given by
\begin{equation}
\frac{d\alpha}{dt} = \Omega_i - \Omega_m - \frac{\alpha}{\tau} \, .
\label{eq:dtalpha}
\end{equation}

We assume that $\Gamma_{icb}$ and $\Gamma_{cmb}$ are both dominated by electromagnetic (EM) forces, and expressed as \citep[e.g.][]{buffett98}
\begin{subequations}
\begin{align}
\Gamma_{cmb} & =  \int_0^{r_f} {\cal F}_m(s) \left[ \Omega_f - \Omega_m \right]   ds  \, , \label{eq:tqcmb} \\
\Gamma_{icb} & =   \int_0^{r_i}  {\cal F}_i(s) \left[ \Omega_f  - \Omega_i \right]  ds \, ,
\label{eq:tqicb}
\end{align}  
\end{subequations}
where the coupling parameters ${\cal F}_m(s)$ and ${\cal F}_i(s)$ depend on the morphology of the magnetic field and the electrical conductivity of the solid side of the boundary.  These are given by
\begin{subequations}
\begin{align}
{\cal F}_m (s) &= 2 \pi s^3 G_m \left(  \frac{z_f}{r_f}  \left( B_{m}^d  \right)^2 + \frac{r_f}{z_f} \langle B_{r,m} \rangle^2 \right)   \, , \label{eq:fms}\\
{\cal F}_i(s) &= 4 \pi s^3 G_i  \frac{z_i}{r_i}  \left( B_{i}^d \right)^2    \, .\label{eq:fmi}
\end{align}
\end{subequations}
where $B_m^d$ and $B_i^d$ are the polar amplitudes of the axial dipole at the CMB and ICB, respectively, $\langle B_{r,m} \rangle$ is the root mean squared (rms) strength of the total radial field (including the axial dipole) at the CMB, $G_m$ is the conductance of the lowermost mantle and $G_i$ is a conductance factor at the ICB given by 
\begin{equation}
G_i = \frac{1}{4} \lbrack 1+ i \,\mbox{sgn}(\omega) \rbrack 
\, \sigma \, \delta \, ,
\end{equation}
where $\sigma$ is the conductivity of the fluid core (assumed equal to that of the inner core), $\delta = \sqrt{2/|\omega| \mu \sigma}$ is the magnetic skin depth at oscillation frequency $\omega$, with $\mu$ the magnetic permeability of free space, and $\mbox{sgn}(\omega) = \omega/|\omega|$.  The form of ${\cal F}_m(s)$ was shown in \citet{dumberry08c} to be a good match with the EM coupling resulting from a realistic field model at the CMB comprised of a large number of spherical harmonic terms.  The field at the ICB is undoubtedly more complex than a simple axial dipole, although the latter is expected to be its dominant contribution, and it is sufficient to capture the first order effect of EM coupling at the ICB.  

\subsection{The MICG mode}

To illustrate how the MICG mode emerges from the above system, let us assume a perfectly rigid inner core ($\tau = \infty$), and no coupling at the CMB (${\cal F}_m(s) =0$).  Let us further assume that the fluid in the ROTC is not involved and remains at rest, while the fluid inside the TC rotates as a single rigid body with angular velocity $\Omega_f^{tc}$.   Taking the time-derivatives of eqs. \eqref{eq:tqm}, \eqref{eq:tqi} and \eqref{eq:tqf}, and using eq. \eqref{eq:dtalpha}, gives
\begin{subequations}
\begin{align}
C_m \frac{d^2  \Omega_m}{dt^2} & =  \overline\Gamma (\Omega_i -  \Omega_m) \, , \label{eq:tqm2} \\
C_i  \frac{d^2 \Omega_i}{dt^2}  & = -\overline\Gamma (\Omega_i - \Omega_m) + r_i \,\overline{\cal F}_i \frac{d}{dt}  (\Omega_f^{tc} - \Omega_i)  \, , \label{eq:tqi2} \\
C_f^{tc}  \frac{d^2 \Omega_f^{tc}}{dt^2}  & =  -  r_i \, \overline{\cal F}_i \frac{d}{dt}  (\Omega_f^{tc} - \Omega_i) \, , \label{eq:tqf2}
\end{align}
\label{eq:micgsys}
\end{subequations}
where $C_f^{tc}$ is the moment of inertia of the fluid inside the TC and $\overline{\cal F}_i$ is the integrated average of  ${\cal F}_i(s)$ over the ICB.  This forms a simple system of axial oscillations between $\Omega_m$, $\Omega_i$ and $\Omega_f^{tc}$, with a restoring gravitational torque between the inner core and mantle and a restoring (and dissipating) EM torque between the inner core and the fluid inside the TC.   The natural modes of oscillations of this system include the MICG mode, whose frequency depends on the strength of $\overline{\Gamma}$ and $\overline{\cal F}_i$.  

Let us consider two end-member scenarios for $\overline{\cal F}_i$.  For no EM coupling at the ICB ($\overline{\cal F}_i=0$), the fluid inside the TC is decoupled and not involved in the MICG mode.  The frequency and eigenvector of the MICG mode are \citep[e.g.][]{dumberry10b}, 
\begin{equation}
\omega_{micg} = \sqrt{\frac{\bar\Gamma(C_m+C_i)}{C_mC_i}}  \, , \hspace*{0.6cm} \left[\Omega_m, \Omega_i \right] = \left[1, -\frac{C_m}{C_i} \right] \, .\label{eq:micg1}
\end{equation}
The oscillation (and angular momentum exchange) is between the mantle and inner core, with no involvement  of the fluid core.  As an example, taking $\overline{\Gamma} = 3 \times10^{20}$ N m, and $C_m$ and $C_i$ given in Table 1, this gives a MICG period of 2.8 years.  If instead $\overline{\cal F}_i$ is very large, the differential rotation between $\Omega_f^{tc}$ and $\Omega_i$ must remain very small (over the period of the MICG mode), and $\Omega_f^{tc}$ and $\Omega_i$ are locked into a common rotation.  Substituting the EM coupling term of eq. \eqref{eq:tqf2} into eq. \eqref{eq:tqi2}, and assuming $\Omega_f^{tc} \approx \Omega_i$, the frequency and eigenvector of the MICG mode are now
\begin{equation}
\omega_{micg} = \sqrt{\frac{\bar\Gamma(C_m+C_{tc})}{C_m C_{tc}}}  \, , \hspace*{0.6cm} \left[\Omega_m, \Omega_i \right] = \left[1, -\frac{C_m}{C_{tc}} \right]  \, ,\label{eq:micg2}
\end{equation}
where $C_{tc}=C_i + C_f^{tc}$.  Even though the gravitational torque remains between the mantle and inner core, the oscillation in this case is between the mantle and the whole of the TC (including the inner core). The larger moment of inertia of the whole of the TC, compared to the inner core alone, reduces the frequency of the MICG and lengthens its period. Using $\overline{\Gamma} = 3 \times10^{20}$ N m as above, the period is now 5.9 years, approximately twice as long.  If $\overline{\cal F}_i$ is somewhere between these two extremes, with an amplitude such that the time it takes for $\Omega_i$ and $\Omega_f^{tc}$ to be brought back into co-rotation is of the order a few years, then the MICG mode is more complex, with each of $\Omega_m$, $\Omega_i$ and $\Omega_f^{tc}$ oscillating at different phases.

At periods of a few years, and assuming a radial magnetic field strength at the ICB of the order of 1 mT or larger, EM coupling is sufficiently strong to prevent any large differential rotation at the ICB \citep{gubbins81}. A rotation of the inner core will tend to entrain the fluid in contact with the ICB into co-rotation.  This zonal motion  is communicated by the (strong) Coriolis force to the whole of the fluid column above and below the inner core, forcing it to rotate rigidly with it in accordance with the Proudman-Taylor constraint \citep{jault08}. Hence, on dynamical grounds, and for the expected strength of the magnetic field at the ICB, we expect that the MICG mode involves the whole of the TC, not just the inner core.        

This simple presentation of the MICG mode is idealized in several ways. First, although $\overline{\cal F}_i$ is most likely strong enough to lock the fluid cylinders inside the TC to the rotational motion of the inner core, some degree of differential rotation can occur near the outer edge of the TC  \citep[e.g.][]{buffett96b,buffett98}.  Second, we have neglected the global elastic deformations that occur in response to the rotational displacement of the inner core out of its equilibrium longitudinal alignment.  Taking these into account would lengthen the MICG period, though not by a large fraction.  Third, if the inner core relaxes viscously over a relatively short time (approaching the MICG period), this would attenuate the mode significantly. 

Perhaps the most important dynamical omission from this idealized picture though is the role played by the fluid in the ROTC.  The latter has been considered in a number of studies \citep[e.g.][]{mound03,mound05b,buffett05,mound06}, although under the assumption that the magnetic field permeating the entire core is relatively weak, with a strength of the order of 0.3 to 0.4 mT.  The primary motivation for our study is precisely to reassess how the dynamics of the ROTC affects the MICG mode when the magnetic field inside the core is much stronger.

\subsection{Torsional oscillation modes}
 
The dynamical equation that governs the evolution of $\Omega_f $ is obtained by integrating the azimuthal component of the momentum equation over a cylindrical surface in the fluid core.  For this specific geometry, the Coriolis acceleration term vanishes \citep[e.g.][]{taylor63,braginsky70}, giving
\begin{equation}
\begin{split}
c_f \frac{\partial \Omega_f}{\partial t} & = \frac{1}{\rho \mu} \frac{\partial }{\partial s} \left(c_f \{B_s\} \frac{b_\phi}{s} \right)  - {\cal F}_m \left[ \Omega_f - \Omega_m \right] \\
&\hspace*{1cm} - {\cal F}_i \left[ \Omega_f - \Omega_i \right]  + \nu \frac{\partial}{\partial s} \left(c_f \frac{\partial \Omega_f}{\partial s} \right) \, , \label{eq:omf} 
\end{split}
\end{equation}
where $\nu$ is the kinematic viscosity, $\{B_s\}$ is the rms strength of the $s$-component of the background magnetic field averaged over a cylindrical surface, and $b_\phi$ is the azimuthal magnetic field perturbation induced by the differential motion of the cylinders. The latter obeys the induction equation 
\begin{equation}
\frac{\partial b_\phi}{\partial t}  = s \{B_s\} \frac{\partial \Omega_f}{\partial s} + \eta \left(  \frac{\partial^2 b_\phi}{\partial s^2}  + \frac{1}{s} \frac{\partial b_\phi}{\partial s} - \frac{b_\phi}{s^2} \right) \, , \label{eq:bphi}
\end{equation}
where $\eta=1/\mu \sigma$ is the magnetic diffusivity.  

When both viscous and magnetic diffusion are neglected, inserting eq. \eqref{eq:bphi} into eq. \eqref{eq:omf}, we retrieve the characteristic equation for Alfv\'en waves subject to EM friction at the top and bottom of the cylinders, \citep[e.g.][]{braginsky70,buffett98}
\begin{equation}
\begin{split}
c_f \frac{\partial^2 \Omega_f}{\partial t^2}  &= \frac{1}{\rho \mu} \frac{\partial }{\partial s} \left( c_f \{ B_s^2 \} \frac{\partial \Omega_f}{\partial s} \right)  \\
& \hspace*{0.5cm}- {\cal F}_m \frac{\partial}{\partial t} \left[ \Omega_f - \Omega_m \right] - {\cal F}_i \frac{\partial}{\partial t} \left[ \Omega_f - \Omega_i \right]  \, ,\label{eq:to} 
\end{split}
\end{equation}
where we have assumed $\{ B_s\}^2 \approx \{ B_s^2 \}$. The normal modes of oscillations of $\Omega_f$ that obey eq. \eqref{eq:to} are the torsional oscillations (TO) modes \citep[e.g.][]{braginsky70}.  In their simplest form, when assuming ${\cal F}_m={\cal F}_i=0$, these modes describe the set of azimuthal oscillations of cylinders maintained by the magnetic tension between them.  The terms that contain ${\cal F}_m$ and ${\cal F}_i$ capture the coupling of the fluid cylinders with the mantle (at the CMB) and the inner core (at the ICB).  Keeping these terms affects the structure of the modes and provides a way by which TO modes can exchange angular momentum with the mantle and inner core.

In most previous studies on TO modes coupled with the mantle and inner core, eq. \eqref{eq:to} was assumed for the dynamics of $\Omega_f$ \citep{buffett96b,buffett98,mound03,mound05b,dumberry08c,dumberry10b}. The studies of \citet{braginsky84} and \citet{mound07} included the effects of viscous diffusion both within the volume of the fluid core and at boundaries.  Here, we retain both magnetic and viscous diffusion and solve for both $\Omega_f$ and $b_\phi$ using the coupled eqs. \eqref{eq:omf} and \eqref{eq:bphi}.  Although we expect $\nu \ll \eta$ in the fluid core, the motivation for retaining viscous diffusion in \eqref{eq:omf} is for numerical convenience, as it makes it easier to prescribe appropriate boundary conditions for $\Omega_f$.  For all our results, we have used $\nu=\eta/100$, so as to ensure that magnetic diffusion dominates over viscous diffusion.  Although it would be preferable to solve for $\Omega_f$ and $b_\phi$ as a function of both $s$ and $z$ \citep[e.g.][]{luo22a}, our simple one-dimensional (1D) model captures the leading order dynamics of Alfv\'en waves within the core.

As we shall show, whether the MICG mode is an independent mode of the core-mantle system depends on the travel time versus diffusion time of Alfv\'en waves in the ROTC.  Alfv\'en waves travel at velocity $V_{A} = \{ B_s \} /\sqrt{\rho \mu}$, and the time they take to travel between the TC and the equator of the CMB ($s=r_f$) is $\tau_A =(r_f-r_i)/ V_A$.  Their typical timescale of diffusion from ohmic dissipation within the core is $\tau_d = 1/(k_s^2 \eta)$, where $k_s$ is the wavenumber in the $s$-direction.   Alfv\'en waves obey the dispersion relation $\omega = k_s V_A$, so TO modes that have frequencies close to MICG mode should have a typical wavenumber $k_s = \omega_{micg}/V_A$, giving $\tau_d = V_A^2 / \eta \omega_{micg}^2$.  Alfv\'en waves are also attenuated by EM coupling at the CMB.  A simple estimate of this EM attenuation timescale can be obtained from eq. \eqref{eq:omf} and is $\tau_{em} = c_f/{\cal F}_m$.  This quantity depends on $s$ but a rough bulk estimate is given by $\tau_{em} = \rho r_c / (G_m \langle B_{r,m} \rangle^2)$.  The travel time versus diffusion time of Alfv\'en waves is then characterized by two Lundquist numbers, 
\begin{equation}
Lu = \frac{\tau_d}{\tau_A} \, , \qquad Lu_{em} =  \frac{\tau_{em}}{\tau_A} \, .
\end{equation}
If either $Lu<1$ or $Lu_{em}<1$, Alfv\'en waves are attenuated before they travel the full width of the ROTC.

Another useful diagnostic quantity is the Lundquist number based on the travel versus diffusion time of Alfv\'en waves in the conducting mantle, 
  \begin{equation}
Lu_m = \sqrt{\frac{\mu}{\rho}} \, G_m \langle B_{r,m} \rangle \, .
\end{equation}
In particular, $Lu_m$ determines the degree of reflection of Alfv\'en waves when they reach the equator of the CMB; when $Lu_m \approx 1$, Alfv\'en waves are fully absorbed \citep{schaeffer16,gillet17}.  This quantity is denoted as $Q$ in both \citet{schaeffer16} and \citet{gillet17}, but here we reserve $Q$ to denote the quality factor of modes.

\section{Numerical implementation}

Eqs. \eqref{eq:omf} and \eqref{eq:bphi} are discretized on a set of grid points in the $s$-direction, with derivatives in $s$ approximated by second-order finite differences. The points $s=r_i=s_{tc}$ and $s=r_f$ are singular in eq. \eqref{eq:omf} because $d c_f/ds$ diverges at these locations.  Likewise, the point $s=0$ is singular in eq. \eqref{eq:bphi}. To avoid these numerical difficulties, we set the innermost and outermost grid points at $s_o=0.01$ and $s_e=0.999$, respectively, and modify the definition of $z_i$ in the vicinity of $s=s_{tc}$ to 
\begin{equation}
\tilde{z}_i = {z_i} - {z_i} \left[ 1 - \left(\frac{s-s_{tc}}{\zeta}\right) \right] \exp \left(\frac{s-s_{tc}}{\zeta}\right)
\, ,\label{eq:zitilde}
\end{equation}
where $\zeta\ll1$ is an adjustable parameter.  With this new definition, $d\tilde{z}_i/ds = 0$ at $s=s_{tc}$, removing the discontinuity at the TC.  The smaller $\zeta$ is, the more confined the modification is to the region close to the TC.  For all our results, we used $\zeta=0.001$.  

Eqs. \eqref{eq:omf} and \eqref{eq:bphi} each require boundary conditions at $s_o$ and $s_e$. At $s=s_o$, we impose the regularity conditions $d\Omega_f/ds=0$ and $b_\phi=0$.  At $s=s_e$, the condition on $b_\phi$ is set by the magnetic field perturbation resulting from EM coupling \citep[e.g. eq. 12 of][]{buffett98}
\begin{equation}
b_\phi(s_e) = -\langle B_{r,m} \rangle  \mu G_m s_e \left(\Omega_f(s_e) - \Omega_m \right) \, ,
\end{equation}
and we impose a free-slip condition, $d\Omega_f/ds=0$.  The latter implies that there is no viscous torque between the fluid core and mantle and minimizes the influence of viscosity on our solutions. Since we use $\nu=\eta/100$, the free-slip condition is met by a flow adjustment in a thin viscous boundary layer near $s=s_e$.

With these boundary conditions, integrating eq. \eqref{eq:omf} between $s_o$ and $s_e$ yields
\begin{equation}
\int_{s_o}^{s_e} c_f \frac{d \Omega_f}{dt}  ds = - \hat{\Gamma}_{cmb} - \hat{\Gamma}_{icb} 
\, , \label{eq:amf2} 
\end{equation}
where $\hat{\Gamma}_{cmb}$ and $\hat{\Gamma}_{icb}$ are modified expressions for the torque on the mantle and inner core given by
\begin{subequations}
\begin{align}
\hat{\Gamma}_{cmb} & =  \int_{s_o}^{s_e} {\cal F}_m(s) \left[ \Omega_f - \Omega_m \right]   ds \nonumber\\
& \hspace*{0.5cm}+  4\pi s_e^3 \sqrt{1-s_e^2} \, G_m \langle B_{r,m} \rangle^2 \left(\Omega_f(s_e) - \Omega_m \right) \, , \label{eq:tqcmb2} \\
\hat{\Gamma}_{icb} & =   \int_{s_o}^{s_{tc}}  {\cal F}_i(s) \left[ \Omega_f  - \Omega_i \right]  ds \, .
\label{eq:tqicb}
\end{align}  
\end{subequations}
The same modified expressions are used for the equations of the mantle \eqref{eq:tqm} and inner core \eqref{eq:tqi} to ensure that angular momentum is conserved.

We specify the location of the discrete $s$-grid points on two Chebyshev grids: one between $s_o$ and $s_{tc}$ and a second between $s_{tc}$ and $s_e$.  The crowding of points near $s_o$, $s_{tc}$ and $s_e$ helps to resolve the thin viscous and magnetic boundary layers at $s_o$ and $s_e$ and to handle the rapidly changing $c_f$ in the region near the TC.  We use 350 points inside the TC and 650 in the ROTC, for a total of $N_s$=1000 grid points, which proved to be sufficient to resolve the structure of the relevant TO modes, and to ensure that angular momentum is conserved to within approximately 1 part in $10^4$. 

The set of eqs. \eqref{eq:tqm}, \eqref{eq:tqi}, \eqref{eq:dtalpha} and the discretized set of eqs. \eqref{eq:omf} and \eqref{eq:bphi} on each of the $N_s$ grid points form a homogeneous system of equations for the set of $n=2N_s + 3$ variables $\Omega_m$, $\Omega_i$, $\alpha$, $\Omega_f(s_k)$ and $b_\phi(s_k)$, where $k=1,2 ... N_s$. We assume that each variable has a time-dependency proportional to $\exp(-i \omega t)$, where $\omega$ is a complex frequency.  The system of equations can be written in matrix form as an eigenvalue problem with eigenvalues $\omega$ and eigenvectors ${\bf x}$,
\begin{equation}
\omega {\mathsf B}\cdot {\bf x} = {\mathsf A}\cdot {\bf x} \, , \label{eq:eigen}
\end{equation}
where ${\mathsf A}$ and ${\mathsf B}$ are $n \times n$ matrices. The solutions are the possible free modes of the system. The imaginary part of $\omega$ is negative, capturing mode decay.  The real part of $\omega$ is either positive or negative (with the same magnitude) and represents the frequency of oscillation; we only solve for the positive frequencies.  Note that ${\cal F}_i$ in eq. \eqref{eq:fmi} has a weak dependence on $\omega$ through the magnetic skin depth $\delta$. However, for a radial magnetic field strength at the ICB $\ge 1$ mT, the fluid cylinders inside the TC become quasi-locked to the inner core, and this frequency dependence only has a very small effect on the normal modes.  To simplify, we  use a nominal frequency of $2\pi/6$ yr$^{-1}$ for calculating $\delta$, such that ${\cal F}_i$ is no longer a function of $\omega$, allowing us to write the system in the simple form of eq. \eqref{eq:eigen}.  The same strategy has been employed in previous studies \citep[e.g.][]{buffett98,buffett05}.

Solutions depend on the choice of the profile of $\{B_s\}$ inside the core. We use a simple prescription given by
\begin{equation}
\{ B_s \} = \overline{B}_s \sin \left(\frac{\pi s}{r_f} \right) + \left(\frac{s}{r_f} \right) B_{s_e}\, ,
\end{equation} 
where $\overline{B}_s$ is an amplitude factor and $B_{s_e}$ is the amplitude of the radial magnetic field at the equator of the CMB.  The sine term produces a maximum $B_s$-field in the middle of the core, similar to the profile inverted by \citet{gillet10}.  The linear term ensures that the radial field at the equator does not vanish.  For our numerical experiments in the weak-field regime, $\overline{B}_s$ is varied between 0.1-0.3 mT, and we set $B_{s_e} = \overline{B}_s/6$.  In the strong-field regime, we set $\overline{B}_s=3.3$ mT and $B_{s_e} = \langle B_{r,m} \rangle$. 

The numerical values that we adopt for all other parameters are given in Table \ref{tab:params}.  The dipole field amplitude $B_m^d=0.319$ mT and the total rms strength of the radial field $\langle B_{r,m} \rangle = 0.391$ mT at the CMB were obtained by taking the time-average of the {\em gufm1} model of \citet{jackson00} over the time period 1950–1990 (the rms of the radial part of the dipole field alone is 0.226 mT). We note that, for our experiments in the weak-field regime, there is an inconsistency between our chosen ICB and CMB field strengths and the lower ${B}_s$-field in the range of 0.1-0.3 mT. These low values were chosen because they are similar to those used in previous studies of the MICG mode \citep[e.g.][]{mound03,mound06} and they help to illustrate the role that Alfv\'en waves have on this mode.

We show results for different choices of mantle conductance: $G_m=10^8$ S,  $2 \times 10^8$ S, $4 \times 10^8$ S, and $6 \times 10^8$ S.  The corresponding $Lu_m$ values, in increasing order of $G_m$, are: 0.34, 0.68, 1.36, 2.05. These remain sufficiently low that our 1D model captures correctly the reflection properties of Alfv\'en waves as they reach the equator of the CMB \citep{schaeffer16,gillet17}.

\begin{table}[h]
\caption{Parameters used in calculations. \label{tab:para}} \vspace*{0.5cm}
\label{tab:params}
\begin{tabular}{@{}ll} \hline
Parameter & value \\
\hline
radius of the core & $r_f = 3.480 \times 10^6$ m \\
radius of the inner core & $r_i = 1.221 \times 10^6$ m \\
moment of inertia, mantle & $C_m = 7.13 \times 10^{37}$ kg m$^{2}$ \\
moment of inertia, inner core & $C_i = 5.87 \times 10^{34}$ kg m$^{2}$ \\
moment of inertia, fluid core & $C_f = 9.42 \times 10^{36}$ kg m$^{2}$ (*)\\
moment of inertia, TC & $C_{tc} = 2.65 \times 10^{35}$ kg m$^{2}$ (*) \\
uniform density, fluid core & $\rho = 1.1 \times 10^4$ kg m$^{-3}$ \\
conductivity, core & $\sigma = 10^6$ S m$^{-1}$ \\
magnetic diffusivity, core & $\eta = \frac{1}{\mu \sigma} = 0.796$ m$^2$ s$^{-1}$ \\
kinematic viscosity, fluid core & $\nu = \frac{\eta}{100} = 0.00796$ m$^2$ s$^{-1}$ \\
rms radial magnetic field, CMB & $\langle B_{r,m} \rangle = 0.319$ mT \\
axial dipole, CMB & $B_m^d = 0.391$ mT \\
axial dipole, ICB & $B_i^d = 3.0$ mT \\
\hline 
\end{tabular} \\
(*) Based on a uniform fluid core density of $\rho = 1.1 \times 10^4$ kg m$^{-3}$.
\end{table}

\section{Results}

Let us first illustrate a case where the modes of axial oscillations between the mantle, inner core and fluid core include a distinct MICG mode. We set $\overline{\Gamma}=3 \times 10^{20}$ N m, which gives an expected MICG frequency of $\omega_{micg} =1.065$ yr$^{-1}$ (period of 5.902 years) based on eq. \eqref{eq:micg2}, and we set the mantle conductance at $G_m=10^8$ S.  We set $\tau=100$ years to ensure that the MICG mode is only weakly damped by the viscous relaxation of the inner core.  We choose $\overline{B}_s=0.1$ mT, which gives a typical travel time for Alfv\'en waves across the ROTC of $\tau_A = 84$ yr.  We compute the set of normal modes of oscillations for this setup, which we refer to as case $a$.

\begin{figure*}[ht!]
  \centering
  \includegraphics[width=0.9\textwidth]{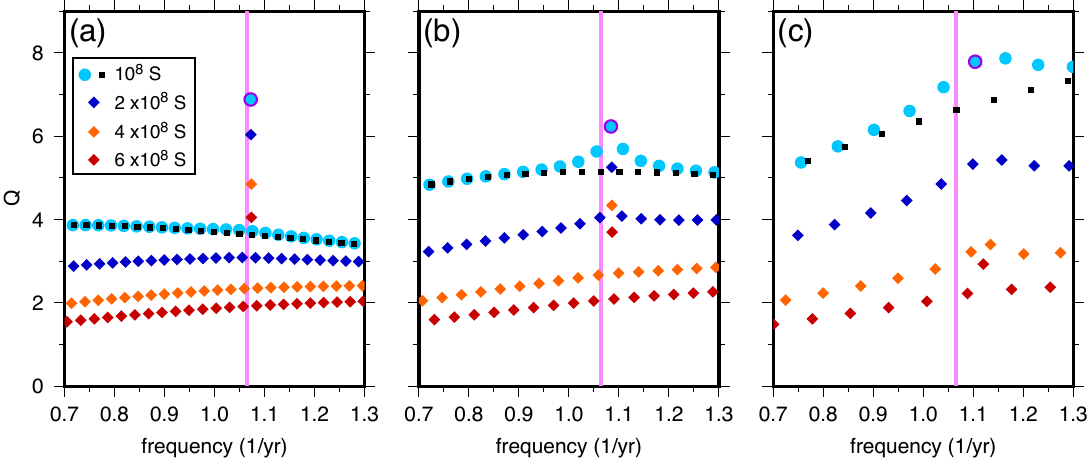} 
  \caption{Frequency and $Q$ of a subset of the normal modes obtained with $\overline{\Gamma}=3 \times10^{20}$ N m, (a) $\overline{B}_s=0.1$, (b) $\overline{B}_s=0.15$, and (c) $\overline{B}_s=0.3$, and different choices of mantle conductance $G_m$ (see legend). Black squares show the TO modes in the absence of gravitational coupling: $\overline{\Gamma}=0$.  In each panel, the predicted $\omega_{micg}$ from Eq. \eqref{eq:micg2} is indicated by the pink vertical line.  The angular velocity structure of the modes identified by a purple outline is shown in Figure \ref{fig:omegaf}.}
  \label{fig:wQ}
\end{figure*}

\begin{figure}[ht!]
  \includegraphics[width=8.4cm]{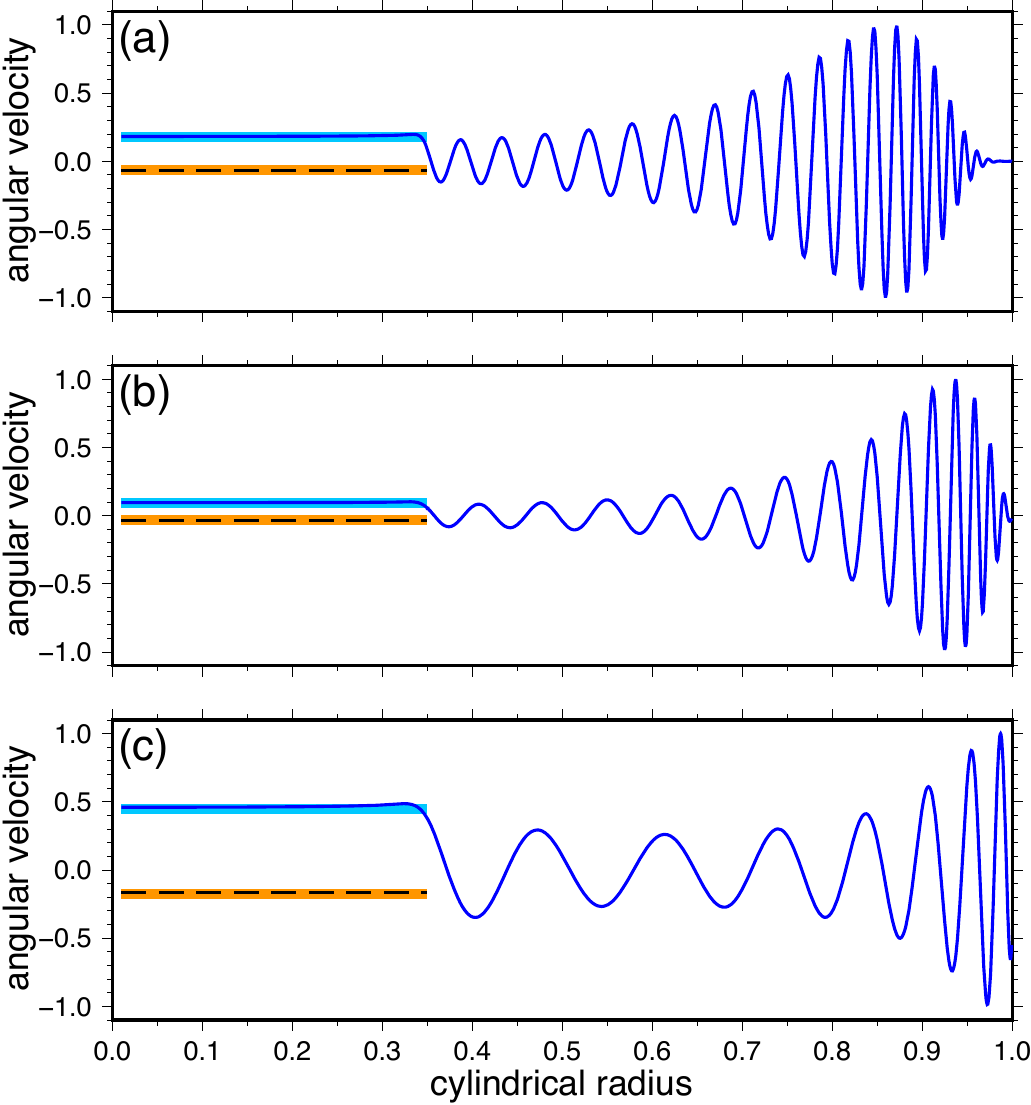} 
  \caption{The angular velocity of the inner core $\Omega_i$ (thick light blue), mantle $\Omega_m \times 100$ (thick orange) and of the fluid core $\Omega_f$ (thin dark blue) as a function of cylindrical radius ($s/r_f$) for the modes identified by a purple outline on the respective panels of Figure \ref{fig:wQ}. The dashed black line corresponds to the prediction $\Omega_m = - (C_{tc}/C_m) \Omega_i$ (multiplied by 100) when the exchange of angular momentum is between the mantle and whole of TC. In each panel, the phase is chosen when $\Omega_i$ is maximum in the prograde direction and the amplitude is normalized such that $\max | \Omega_f |=1$.} 
    \label{fig:omegaf}
\end{figure}

Figure \ref{fig:wQ}a shows the frequency and quality factor $Q$ of a subset of the resulting modes (light blue circles) that have frequencies in the vicinity of $\omega_{micg}$.  For this choice of  $\overline{B}_s$, the fundamental mode (first harmonic) of TO has a period of $\sim180$ years; the sequence of modes that have a $Q$ between 3 and 4 on Figure \ref{fig:wQ}a correspond to higher harmonics of TO.  In addition to these, a distinct mode (identified by a purple outline) with a higher $Q$ of approximately 7, separated from the TO branch, is also present: this is the MICG mode.  Its frequency matches well the expected theoretical prediction under the assumption that the whole of the TC co-rotates with the inner core.  Figure \ref{fig:omegaf}a shows the profile of $\Omega_f$ for this MICG mode, at the phase when $\Omega_i$ is maximum in the prograde direction. Inside the TC, $\Omega_f$ is locked with $\Omega_i$ except near the edge of the TC, as found in  previous studies \citep{buffett96a,buffett98}. The rotation of the mantle, $\Omega_m$, is opposite that of the inner core, and its amplitude matches that  of the MICG mode under the assumption that the whole of the TC oscillates as a rigid body, $\Omega_m = - (C_{tc}/C_m) \Omega_i$. (Note that $\Omega_m$ is increased by a factor 100 on Figure \ref{fig:omegaf} so that it reaches a similar scale as $\Omega_i$ and $\Omega_f$). The departure of $\Omega_f$ from $\Omega_i$ near the edge of the TC implies a slight reduction in the effective moment of inertia of the TC taking part in the MICG mode and explains why its frequency is slightly higher than its theoretical prediction (see Figure \ref{fig:wQ}a).  Also visible in Figure \ref{fig:omegaf}a are the Alfv\'en waves generated from the TC outward.  With $\overline{B}_s=0.1$ mT, and at the frequency of the MICG mode, their wavelength is sufficiently small that their typical time of attenuation is $\tau_d=25$ yr.  This is shorter than $\tau_A$, so $Lu <1$, and Alfv\'en waves get attenuated before they reach the equator of the CMB ($s=r_f$).  These waves carry angular momentum but their short wavelengths imply a small net angular momentum when integrated over the ROTC, and the latter plays a negligible role in the MICG angular momentum budget.

For comparison, we also show on Figure \ref{fig:wQ}a the modes (black squares) that result when there is no gravitational coupling ($\overline{\Gamma}=0$). In this case, the MICG mode is absent from the spectrum of modes; only TO modes are present, and their frequencies and $Q$ coincide almost identically with those found with $\overline{\Gamma}\neq0$.  These results show that when $B_s$ within the core is sufficiently weak, the MICG mode emerges as an independent mode of the system, along with a set of TO modes, and the latter are only weakly influenced by gravitational coupling.  

For a mantle conductance of $G_m = 10^8$, the typical EM attenuation time of Alfv\'en waves is $\tau_{em}=119$ yr.  We also show on Figure \ref{fig:wQ}a the set of modes for three other choices of $G_m$: $2 \times 10^8$ S ($\tau_{em} = 60$ yr), $4 \times 10^8$ S ($\tau_{em}=30$ yr) and $6 \times 10^8$ S ($\tau_{em} = 20$ yr).   The $Q$ of all modes is reduced because of the enhanced attenuation from EM coupling at the CMB. In each case, the MICG mode is present as an independent mode because $Lu<1$, while $Lu_{em}$ is also $<1$ for the 3 cases with higher mantle conductance.

Let us now increase $B_s$ while keeping  $\overline{\Gamma}=3 \times 10^{20}$ N m and restoring $G_m=10^8$ S. We compute the modes for $\overline{B}_s=0.15$ mT (case $b$) and $\overline{B}_s=0.3$ mT (case $c$).  The results are shown on Figure \ref{fig:wQ} (panels b and c, respectively), where we again show the comparison with the modes computed with $\overline{\Gamma}=0$.   Figures \ref{fig:omegaf}b and \ref{fig:omegaf}c show the profiles of $\Omega_f$ for the modes close to the MICG frequency identified by a purple outline in panels $b$ and $c$ of Figure \ref{fig:wQ}.  With a larger $B_s$-field,  Alfv\'en waves generated at the TC travel faster, their wavelengths at frequencies close to $\omega_{micg}$ is longer (increasing their ohmic diffusion time $\tau_{d}$), so they propagate further before they get attenuated.  The travel and ohmic diffusion times for case $b$ are $\tau_A=56$ yr, $\tau_d=58$ yr and those for case $c$ are $\tau_A= 28$ yr, $\tau_d=235$ yr.  In each case, $Lu>1$ (though barely for case $b$) and (with $\tau_{em}=119$ yr) $Lu_{em}>1$; Alfv\'en waves can travel the width of ROTC before they get attenuated. Consequently, the structure of Alfv\'en waves within the whole of the ROTC must be compatible with their end-condition at $s=r_f$ and they thus have a feedback on the oscillating motion of the TC.  In each case, the MICG mode is no longer a distinct, independent mode of the system.  Instead, it merges into the spectrum of TO modes.

Figures \ref{fig:wQ}b and \ref{fig:wQ}c also show the resulting modes when the mantle conductance is increased to $2 \times 10^8$ S ($\tau_{em} = 60$ yr), $4 \times 10^8$ S ($\tau_{em}=30$ yr) and $6 \times 10^8$ S ($\tau_{em} = 20$ yr).  By progressively decreasing $\tau_{em}$, a point is reached when Alfv\'en waves get attenuated before they can travel the width of the ROTC (that is, $Lu_{em}$ becomes smaller than 1), and a distinct MICG mode re-emerges in the spectrum of modes.  With $\overline{B}_s=0.15$ mT ($\tau_A = 56$ yr), the MICG is almost fully out of the TO branch for $G_m = 2 \times 10^8$ S ($Lu_{em} \sim 1$), and clearly separated for $G_m > 4 \times 10^8$ S . With $\overline{B}_s=0.3$ mT ($\tau_A = 28$ yr), a larger $G_m = 6 \times 10^8$ S is required for $Lu_{em}<1$ and for the MICG mode to split from the TO branch.

The above analysis shows that the MICG only remains an independent mode when Alfv\'en waves launched by the oscillating TC get attenuated before they reach $s=r_f$ (either because $Lu<1$ or $Lu_{em}<1$).  If so, they do not have a feedback on the motion of the TC, and the MICG mode can exist independently from the TO modes.  In contrast, when Alfv\'en waves can travel through the entire width of the ROTC, the motion of the TC must be compatible with the structure of TO modes, and the MICG mode gets absorbed in the spectrum of TO modes.   Alfv\'en waves may be partially or fully absorbed once they reach the equator of the CMB \citep{schaeffer16,gillet17}, but the key factor is whether they can reach this point. 

The $B_s$-field chosen for cases $b$ and $c$ above remains weak (smaller than 0.3 mT).  The TO harmonics at periods close to the MICG mode have short wavelengths and the net angular momentum of the ROTC for these modes remains small.  The angular momentum budget for these TO modes remains primarily an exchange between the mantle and the whole of the TC via gravitational coupling.  Figures \ref{fig:omegaf}b and \ref{fig:omegaf}c show that $\Omega_m$ indeed remains approximately equal to $- (C_{tc}/C_m) \Omega_i$ for these selected modes. Hence, these TO modes share the same angular momentum dynamics as the MICG mode.

Although presented differently, our results are in agreement with those of \citet{mound03}.  In their study, which uses a similar model, they do not compute modes directly.  Rather, they impose a forcing on the inner core, and monitor the response of the mantle.  The frequencies at which the angular motion of the mantle is amplified mark the periods of the normal modes.  With a uniform $B_s$-field of 0.4 mT (similar to our case $c$), and with no EM coupling at the CMB, they find (as we do) a set of TO modes  amplified in the broad vicinity of the MICG frequency (see their Figure 4).  This amplification leaves a broad peak in the LOD at the period of the MICG mode \citep{mound06}.

We now explore how the MICG mode affects the TO modes in a strong $B_s$-field regime. We restore $G_m=10^8$ S and set $\overline{B}_s = 3.3$ mT, typical of the field strength expected inside the core \citep{gillet10}.  With these choices, the fundamental TO mode has a period of approximately 6 years. Alfv\'en waves can readily propagate across the entire ROTC in a short timescale of $\tau_A = 2.5$ yr.  This is much faster than $\tau_{em}$ and also much faster than $\tau_d$ which (for a period of 6 yr) is now close to $3 \times 10^4$ yr.  Hence, both $Lu\gg 1$ and $Lu_{em} \gg1$, and, accordingly, the spectrum of modes should not include an independent MICG mode; we have verified that this is indeed the case for any choices of $\overline{\Gamma}$.  With our chosen mantle conductance, $Lu_{m}=0.34$; this is sufficiently high that Alfv\'en waves are partly absorbed at the equator of the CMB, yet low enough that distinct spectral peaks at the periods of the TO modes would remain visible in the LOD spectrum, compatible with observations \citep{gillet17}.

\begin{figure}[ht!]
  \includegraphics[width=8.4cm]{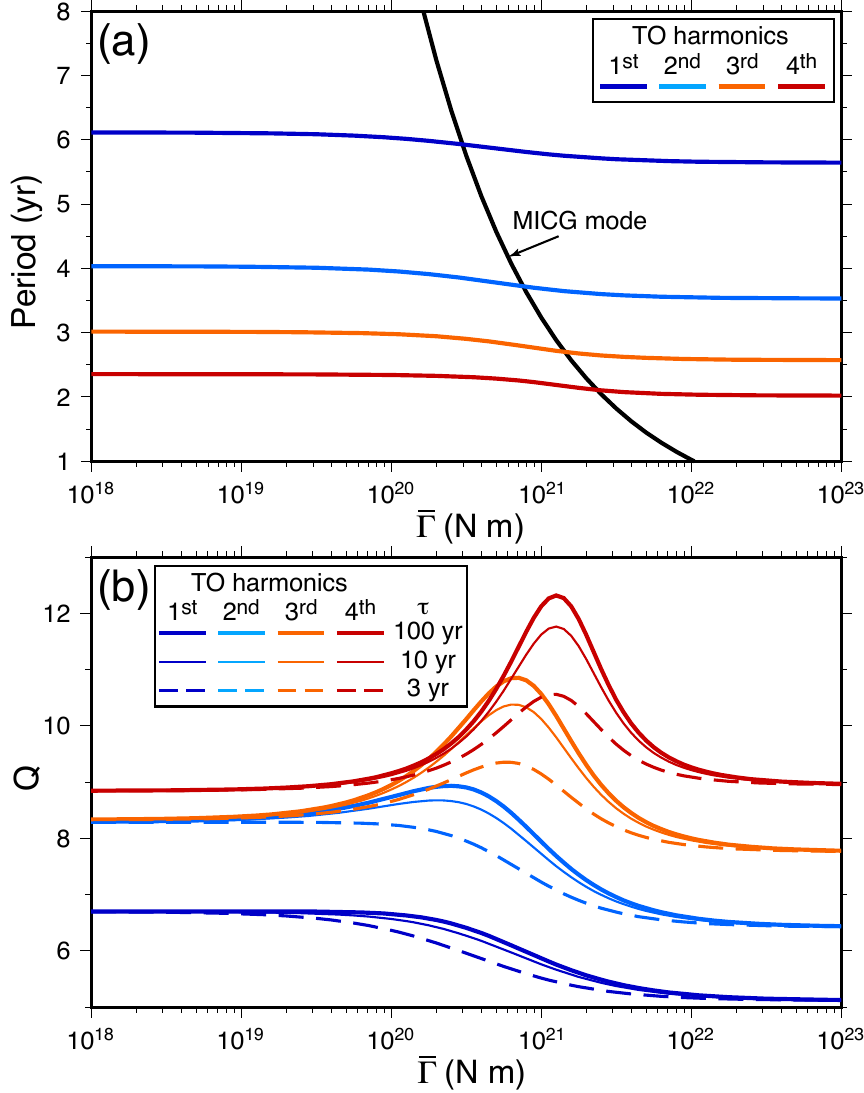} 
  \caption{The period (a) and $Q$ (b) of the first four TO harmonics (coloured lines, legend) as a function of $\overline{\Gamma}$.  The black line in (a) is the period of the MICG mode predicted from Eq. \eqref{eq:micg2}. In (b), results are shown for $\tau=100$ years (solid lines), $\tau=10$ years (thin solid lines) and $\tau=3$ years (dashed lines).}
    \label{fig:TOp}
\end{figure}

\begin{figure*}[ht!]
  \centering
 \includegraphics[width=0.95\textwidth]{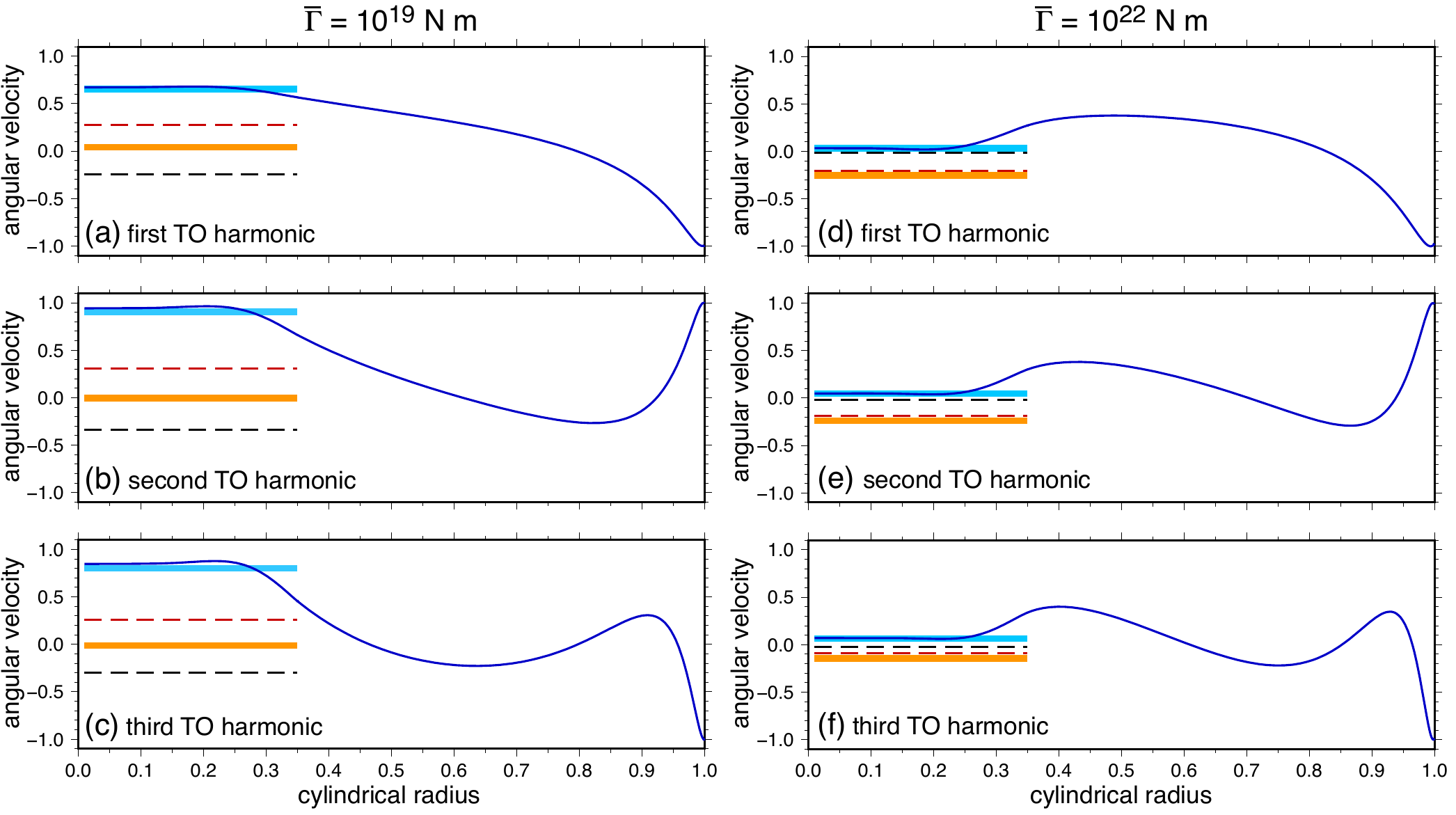} 
 \caption{The angular velocity as a function of cylindrical radius ($s/r_f$) of the (a, d) first, (b, e) second, and (c, f) third TO harmonics for $\overline{\Gamma}=10^{19}$  N m (left column) and $\overline{\Gamma}=10^{22}$  N m (right column).  Thin blue lines: $\Omega_f$, thick light blue lines: $\Omega_i$ and thick orange lines: $\Omega_m \times 100$. The dashed black line corresponds to the prediction $\Omega_m = - (C_{tc}/C_m) \Omega_i$ (multiplied by 100). The dashed red line corresponds to the prediction of $\Omega_m = - L_{rotc}/C_m$ (multiplied by 100), where $L_{rotc}$ is the net angular momentum of the ROTC. In each panel, the phase is chosen when $\Omega_i$ is maximum in the prograde direction and the amplitude is normalized such that $\max | \Omega_f |=1$.} 
    \label{fig:modemorph}
\end{figure*}

Figure \ref{fig:TOp} shows how the periods (panel a) and $Q$ (panel b) of the first four TO harmonics are altered as a function of $\overline{\Gamma}$. Figure \ref{fig:modemorph} shows how the structure of the first three TO harmonics differ in a weak ($\overline{\Gamma} = 10^{19}$ N m) versus a strong ($\overline{\Gamma} = 10^{22}$ N m) gravitational coupling regime. When gravitational coupling is weak ($\overline{\Gamma} = 10^{19}$ N m), the oscillation of the TC -- entrained by the TO modes -- is only weakly resisted by gravitational coupling.  We also show on Figure \ref{fig:modemorph} the predictions of $\Omega_m$ based on an angular momentum exchange with the TC ( $\Omega_m =-(C_{tc}/C_m) \Omega_i$) and with the ROTC ($\Omega_m = - L_{rotc}/C_m$, where $L_{rotc}$ is the net angular momentum of the ROTC).  The sum of these predictions matches $\Omega_m$ well, reflecting that the core-mantle angular momentum is conserved.  But neither provides a good fit to $\Omega_m$, which has a much smaller amplitude.  This indicates that the angular momentum budget for these modes is primarily an exchange between the TC and the ROTC; $\Omega_m$ reflects the small imbalance in the angular momenta of the TC and the ROTC.  When gravitational coupling is strong ($\overline{\Gamma} = 10^{22}$ N m), the oscillation of the TC is more restricted.  Using standing waves on a string as an analogy, a strong gravitational coupling changes the boundary condition on $\Omega_f$ from a free-end at $s\rightarrow0$ to one more akin to a fixed-end at $s<s_{tc}$.  Because the latter implies a shorter mode wavelength, the period of the TO modes are shorter (see Figure \ref{fig:TOp}a).  The angular momentum budget is now dominantly an exchange between the mantle and the ROTC (although the TC plays a non-negligible role).  Whether $\overline{\Gamma}$ is weak or strong, these TO modes have a different angular momentum dynamics than the MICG mode. Instead, the ROTC is involved in the angular momentum budget and it is the whole of the core that exchanges its angular momentum with the mantle. 

Figure \ref{fig:TOp}a also shows that the point of transition between the weak and strong gravitational coupling regime for every TO mode is set by the period of the MICG mode. Hence, even though the MICG mode is no longer an individual mode of the system, its period retains an influence in the characteristics of the resulting TO modes. When the period of a TO mode approaches the MICG period, a resonance effect occurs, leading to a higher $Q$ (Figure \ref{fig:TOp}b).  Note that the level of amplification is larger when the MICG period matches higher TO harmonics, consistent with results obtained in the weak $B_s$-field regime. When the MICG period is longer than the period of the fundamental TO harmonic, the level of amplification is limited. The amplification also depends on the attenuation of the TO modes caused by EM damping at the CMB and by the viscous relaxation time of the inner core $\tau$.  When the latter is of the same order or shorter than the period of a given TO mode, $Q$ is significantly reduced.  This is shown in Figure \ref{fig:TOp}b, where we have computed $Q$ for $\tau=100$, $10$ and $3 $ years.   

\section{Discussion and conclusions}

Our results show that if Alfv\'en waves launched at the TC can travel across the entire width of the ROTC before being attenuated, the MICG mode is no longer a distinct independent mode of the core-mantle system.  For a magnetic field strength in the bulk of the outer core equal to a few mT, as predicted by dynamo models \citep{christensen06,aubert09}, Alfv\'en waves can readily traverse the core in only a few years \citep{gillet10}, much shorter than their attenuation time.  The MICG mode should be absorbed into the spectrum of TO modes.  The MICG period retains a dynamical influence, as it acts as a point of resonance for TO modes, and marks the transition from a TO mode in which the oscillating motion of the TC is weakly affected by gravitational coupling to one in which the motion of the TC is strongly impeded by it.

For the MICG mode to remain an independent mode of the core-mantle system, the attenuation time of Alfv\'en waves by coupling at the CMB (EM or otherwise) would need to be of the order of one year or less.  As an indication, such a fast attenuation requires a mantle conductance larger than $10^9$ S, so large in fact that the gravest TO modes and the MICG mode would all have a $Q$ below 1.   Hence, the very fact that TO modes are excited to an observable level points to a limit in the mantle conductance which is well below $10^9$ S \citep{gillet17}.  In addition, such a large conductance would introduce a significant delay between the time a magnetic signal is generated at the CMB and its observation at the surface \citep[e.g.][]{jault15}, yet no evidence of such a delay is observed \citep[e.g.][]{holme13,gillet15}.  The latest results from mineral physics and seismology also point to an upper bound on the mantle conductance well below $10^9$ S.  The electrical conductivity of FeO at CMB conditions is in the range of $10^4-10^5$ S m$^{-1}$ \citep{ohta12,ho24}, and while a thin layer of approximately 2 km of such iron-enriched material is supported by normal modes \citep{russell23}, a thicker layer of a few tens of km -- such as would be required for a conductance $>10^9$ S -- is not.  Given the limits on the possible mantle conductance, it is difficult to escape the conclusion that the MICG mode should be absorbed in the spectrum of TO modes. 

This conclusion is important regarding the interpretation of the 5.9 year oscillation in the LOD, which includes a clear contribution from the core \citep[e.g.][]{pfeffer23,cazenave25}.  A number of studies have suggested that this signal may represent the signature of the MICG mode \citep{mound06,chao17,ding18,duan18,shih21}.  This interpretation rests on the assumption that the magnetic field in the core is weak (equal to a fraction of a mT). In such a case, the resonant excitation of a cluster of high TO harmonics with periods close to that of the MICG produces a distinct LOD peak at the MICG period \citep{mound03,mound06}.  The angular momentum budget for these TO modes is primarily an exchange between the mantle and the whole of the TC, just as it is for the MICG mode. However, when the magnetic field strength inside the core is a few mT (as we expect), the subdecadal period range is now populated by the first few TO harmonics. These have distinct, separated periods and their angular momentum dynamics feature a leading order contribution from the ROTC; they do not behave like the MICG mode.  These TO modes have a higher $Q$ if their periods are close to the MICG period, but the peaks in the LOD that they may leave correspond to the distinct periods of the TO modes, not the period of the MICG.  Hence, for a strong magnetic field inside the core of a few mT, the 5.9 year LOD signal cannot represent a direct signature of the MICG mode. Instead, it must be a signature of the fundamental TO mode, or any other free or driven core flow motion. This is consistent with the results of \citet{gillet10} who showed that, for the 5.9 year LOD signal, the mantle exchanges its angular momentum with the whole of the core, not the TC alone. 

The most recent estimates of the gravitational coupling constant $\overline{\Gamma}$ based on mantle flow models suggest a range of $0.3 - 2 \times 10^{20}$ N m \citep{davies14}, corresponding to a range of MICG periods between 7 and 19 years.  Assuming a $B_s$-field of approximately 3 mT or larger, all TO modes should then have a period shorter than the MICG period.  This implies that TO modes should be weakly influenced by gravitational coupling in the sense that oscillations of the TC region should not be severely restricted by the mantle.  We thus expect that, at subdecadal periods, the TC (including the inner core) can undergo oscillating motion with respect to the mantle at a similar angular velocity amplitude as cylinders outside the TC. At a period of 6 years, and based on the observed fluctuations of the magnetic field, zonal flow fluctuations inside the TC have an amplitude of approximately $v_\phi \sim0.4$ km yr$^{-1}$ \citep{gillet10}. Using this as a representative measure, this corresponds to longitudinal angular fluctuations of the inner core of approximately $0.02^\circ$ \citep{lecomte23}. 

Our results also have important implications for the interpretation of the subdecadal fluctuations in inner core rotations recently inferred by seismic observations, notably in the study of \citet{wang22} which suggests a super-rotation of the inner core of $\sim0.29^{\circ}$ between 1971 and 1974 and a sub-rotation of $\sim0.1^\circ$ between 1969 and 1971.  \citet{wang22} further showed that the amplitude and phase of these inferred inner core rotation changes ($\Omega_i$) are consistent with those reconstructed from the 5.9 year LOD signal under the assumption that the latter is caused by the MICG mode.  The authors have themselves challenged their own interpretation in a subsequent study focused on the time window 1991-2023, in which a 6-yr inner core oscillation of a similar magnitude is not observed \citep{wang24}. Based on our results, we favour an interpretation of subdecadal changes in $\Omega_i$ entrained by TO modes or, more generally, by Alfv\'en waves.   In this case, unlike for the MICG mode, a prediction of $\Omega_i$ cannot be constructed directly from the LOD signal but instead depends on many factors, including the relative strengths of the EM and gravitational torque on the mantle and the structure of Alfv\'en waves.  It is of course possible that the phase of $\Omega_i$ entrained by Alfv\'en waves is similar to that predicted from the LOD signal on the basis of the MICG mode.  However, the amplitude of the subdecadal changes $\Omega_i$ should be limited to  approximately $0.02^\circ$, as stated above, much smaller than the amplitudes inferred by \citet{wang22}.

Although our focus here has been on the MICG mode, our model may be adapted to monitor the angular momentum response of the core-mantle system to a specified forcing in the core, either at a given frequency \citep[e.g.][]{mound03,mound05b} or for a stochastic forcing \citep[e.g.][]{gillet17}.  The parameter space of $\overline{\Gamma}$, $\tau$ and $G_m$ may be surveyed to determine the range of values that allow to best reproduce the characteristics of the observed Alfv\'en waves, while also matching the amplitude and phase of the LOD changes at subdecadal and decadal timescales.  This would provide constraints on these parameters, which in turn would provide further constraints on the structure, composition and dynamics of the lower mantle and inner core. 

\begin{acknowledgements}
I wish to thank Jon Mound and an anonymous reviewer for comments and suggestions that have greatly improved the quality of this paper.  This work was financially supported by a Discovery Grant from NSERC/CRSNG (Canada).  Figures were produced using the GMT software \citep{gmt}.
\end{acknowledgements}

\section*{Data availability}

Codes to reproduce all numerical experiments are freely accessible at the following data repository: \citet{dumberry25data}.

\section*{Competing interests}  

The author has no competing interests.

\printbibliography

@String{AN = "Astronomische Nachrichten"}

@string{EPSL = {Earth Planet. Sci. Lett.}}

@string{GAFD = {Geophys. Astrophys. Fluid Dyn}}

@string{GJI = {Geophys. J. Int.}}

@string{GRL = {Geophys. Res. Lett.}}

@string{JGR = {J. Geophys. Res.}}

@string{PEPI = {Phys. Earth Planet. Inter.}}

@String{PRL = "Phys. Rev. Lett."}

@String{PTRSLA = "Phil. Trans. R. Soc. Lond. A"}

@String{PRSLA = "Proc. R. Soc. Lond. A"}

@Article{abarcadelrio00,
  author = 	 {Abarca del Rio, R. and Gambis, D. and Salstein, D. A.},
  title = 	 {Interannual signals in length of day and atmospheric angular momentum},
  journal = {Ann. Geophys.},
  year = 	 2000,
  volume =	 18,
  pages =	 {347--364},
  doi = {10.1007/s00585-000-0347-9}
}

@article{aubert09,
  title={Modelling the palaeo-evolution of the geodynamo},
  author={Aubert, Julien and Labrosse, St{\'e}phane and Poitou, Charles},
  journal={Geophys. J. Int.},
  volume={179},
  pages={1414--1428},
  year={2009},
  doi = {10.1111/j.1365-246x.2009.04361.x}
}

@Article{aubert11,
  author = 	 {Aubert, J. and Dumberry, M.},
  title = 	 {Steady and fluctuating inner core rotation in numerical geodynamo models},
  journal = 	 GJI,
  year = 	 2011,
  volume =	 184,
  pages =	 {162--170},
  doi = {10.1111/j.1365-246x.2010.04842.x}
}

@Article{braginsky70,
  author =	 {Braginsky, S. I.},
  title =	 {Torsional magnetohydrodynamic vibrations in the
                  {E}arth's core and variations in day length},
  journal =	 {Geomag. Aeron.},
  year =	 1970,
  volume =	 10,
  pages =	 {1--10}
}

@Article{braginsky84,
  author =	 {Braginsky, S. I.},
  title =	 {Short-period geomagnetic secular variation},
  journal =	 GAFD,
  year =	 1984,
  volume =	 30,
  pages =	 {1--78},
  doi = {10.1080/03091928408210077}
}

@Article{buffett96a,
  author = 	 {Buffett, B. A.},
  title = 	 {Gravitational oscillations in the length of the day},
  journal = 	 GRL,
  year = 	 {1996},
  volume =	 23,
  pages =	 {2279--2282},
  doi = {10.1029/96gl02083}
}

@Article{buffett96b,
  author = 	 {Buffett, B. A.},
  title = 	 {A mechanism for fluctuations in the length of day},
  journal = 	 GRL,
  year = 	 {1996},
  volume =	 23,
  pages =	 {3803--3806},
  doi = {10.1029/96gl03571}
}

@InCollection{buffett98,
  author = 	 {Buffett, B. A.},
  title = 	 {Free oscillations in the length of day: inferences on physical properties near the core-mantle boundary},
  booktitle = 	 {The core-mantle boundary region},
  pages =	 {153--165},
  publisher =	 {{AGU} {G}eophysical {M}onograph },
  year =	 1998,
  editor =	 {Gurnis, M. and Wysession, M. E. and Knittle, E. and Buffett, B. A.},
  volume =	 28,
  series =	 {Geodynamics series},
  address =      {Washington, DC}
}

@Article{buffett00,
  author = 	 {Buffett, B. A. and Glatzmaier, G. A.},
  title = 	 {Gravitational braking of inner-core rotation in geodynamo simulations},
  journal = 	 GRL,
  year = 	 2000,
  volume =	 27,
  pages =	 {3125--3128},
  doi = {10.1029/2000gl011705}
}

@Article{buffett05,
  author = 	 {Buffett, B. A. and Mound, J. E.},
  title = 	 {A {G}reen's function for the excitation of torsional oscillations in the {E}arth's core},
  journal = {J. Geophys. Res. Solid Earth},
  year = 	 2005,
  volume =	 110,
  pages  =         {B08104},
  doi = {10.1029/2004JB003495}
}

@Article{cazenave25,
  author = 	 {Cazenave, A. and Pfeffer, J. and Mandea, M. and Dehant, V. and Gillet, N.},
  title = 	 {Why is the {E}arth system oscillating at a 6-year period?},
  journal = {Surv. Geophys.},
  year = 	 2025,
  volume =	 46,
  pages  =  {503--528},
  doi = {10.1007/s10712-024-09874-4}
}

@Article{chao14,
  author = 	 {Chao, B. F. and Chung, W. and Shih Z. and Hsieh, Y.},
  title = 	 {{E}arth's rotation variations: a wavelet analysis},
  journal = 	 {Terra Nova},
  year = 	 2014,
  volume =	 26,
  pages =	 {260--264},
  doi = {10.1111/ter.12094}
}

@Article{chao17,
  author = 	 {Chao, B. F.},
  title = 	 {Dynamics of axial torsional libration under the mantle-inner
core gravitational interaction},
  journal =  {J. Geophys. Res. Solid Earth},
  year = 	 2017,
  volume =	 122,
  pages =	 {560--571},
  doi = {10.1002/2016jb013515}
}

@Article{christensen06,
  author = 	 {Christensen, U. R. and Aubert, J.},
  title = 	 {Scaling properties of convection-driven dynamos in rotating spherical shells and application to planetary magnetic fields},
  journal = 	 GJI,
  year = 	 2006,
  volume =	 166,
  pages =	 {97--114},
  doi = {10.1111/j.1365-246x.2006.03009.x}
}

@Article{davies14,
  author = 	 {Davies, C. J. and Stegman, D. R. and Dumberry, M.},
  title = 	 {The strength of gravitational core-mantle coupling},
  journal = 	 GRL,
  year = 	 2014,
  volume =	 41,
  pages =	 {3786--3792},
  doi = {10.1002/2014gl059836}
}

@Article{defraigne96,
  author = 	 {Defraigne, P. and Dehant, V. and Wahr, J. M.},
  title = 	 {Internal loading of an inhomogeneous compressible {E}arth with phase boundaries},
  journal = 	 GJI,
  year = 	 1996,
  volume =	 125,
  pages =	 {173--192},
  doi = {10.1111/j.1365-246x.1996.tb06544.x}
}

@Article{ding18,
  author = 	 {Ding, H. and Chao, B. F.},
  title = 	 {A 6-yr westward rotary motion in the {E}arth: {D}etection and possible {MICG} coupling mechanism},
  journal = 	 EPSL,
  year = 	 2018,
  volume =	 495,
  pages =	 {50--55},
  doi = {10.1016/j.epsl.2018.05.009}
}

@Article{duan18,
  author = 	 {Duan, P. and Liu, G. and Hu, X. and Zhao, J. and Huang, C.},
  title = 	 {Mechanism of the interannual oscillations in length of day and its constraint on the electromagnetic coupling at the core-mantle boundary},
  journal = EPSL,
  year = 	 2018,
  volume =	 482,
  pages =	 {245--252},
  doi = {10.1016/j.epsl.2017.11.007}
}

@Article{dumberry08c,
  author = 	 {Dumberry, M. and Mound, J. E.},
  title = 	 {Constraints on core-mantle electromagnetic coupling from torsional oscillation normal modes}, 
  year = 	 2008,
  journal =  {J. Geophys. Res. Solid Earth},
  volume =	 113,
  pages =	 {B03102},
  doi = {10.1029/2007jb005135}
}

@Article{dumberry10b,
  author = 	 {Dumberry, M. and Mound, J. E.},
  title = 	 {Inner core -- mantle gravitational locking and the super-rotation of the inner core}, 
  year = 	 2010,
  journal =      GJI,
  volume =	 181,
  pages =	 {806--817},
  doi = {10.1111/j.1365-246x.2010.04563.x}
}

@data{dumberry25data,
  author = 	 {Dumberry, M.},
  title = 	 {Replication Data for: The mantle-inner core gravitational mode of oscillation in a strong magnetic field regime}, 
  year = 	 2025,
  publisher  = {Borealis},
  version = {V2},
  doi = {10.5683/SP3/HF3RJQ},
}

@Article{forte94,
  author = 	 {Forte, A. M. and Woodward, R. L. and Dziewonski, A.M.},
  title = 	 {Joint inversions of seismic and geodynamic data for models of three-dimensional mantle heterogeneity},
  journal =  {J. Geophys. Res. Solid Earth},
  year = 	 1994,
  volume = 99,
  pages = 	 {21857--21877},
  doi = {10.1029/94jb01467}
}

@article{garnero16,
  author = 	 {Garnero, E. J. and McNamara, A. K. and Shim, S.-H.},
  title = 	 {Continent-sized anomalous zones with low seismic velocity at the base of {E}arth's mantle},
  journal = 	 {Nat. Geosci.},
  year = 	 2016,
  volume =	 9,
  pages =	 {481--489},
  doi = {10.1038/ngeo2733}
}

@article{gillet10,
  author =       {Gillet, N. and Jault, D. and Canet, E. and Fournier, A.},
  journal =      {Nature},
  title =        {Fast torsional waves and strong magnetic field within the {E}arth's core},
  year = 	 2010,
  volume =	 465,
  pages =	 {74--77},
  doi = {10.1038/nature09010}
}

@article{gillet15,
  author =       {Gillet, N. and Jault, D. and Finlay, C. C.},
  journal =      {J. Geophys. Res. Solid Earth},
  title =        {Planetary gyre, time-dependent eddies, torsional waves,
and equatorial jets at the {E}arth's core surface},
  year = 	 2015,
  volume =	 120,
  pages =	 {3991--4013},
  doi = {10.1002/2014jb011786}
}

@article{gillet17,
  author =       {Gillet, N. and Jault, D. and Canet, E.},
  journal =      GJI,
  title =        {Excitation of traveling torsional normal modes in an {E}arth's core model},
  year = 	 2017,
  volume =	 210,
  pages =	 {1503--1516},
  doi = {10.1093/gji/ggx237}
}

@Article{gmt,
  author = 	 {Wessel, P. and Smith, W. H. F. and Scharroo, R. and Luis, J. and Wobbe, F.},
  title = 	 {{Generic Mapping Tools}: Improved Version Released},
  journal =  {EOS Trans. AGU},
  year = 	 2013,
  volume =	 94,
  pages =	 {409--410},
  doi = {10.1002/2013EO450001}
}

@Article{gubbins81,
  author = 	 {Gubbins, D.},
  title = 	 {Rotation of the inner core},
  journal = 	 JGR,
  year = 	 1981,
  volume =	 86,
  pages =	 {11695--11699},
  doi = {10.1029/jb086ib12p11695}
}

@Article{hager85,
  author = 	 {Hager, B. H. and Clayton, R. W. and Richards, M. A. and Comer, R. P. and Dziewonski, A.},
  title = 	 {Lower mantle heterogeneity, dynamic topography and the geoid},
  journal = 	 {Nature},
  year = 	 1985,
  volume = 	 313,
  pages = 	 {541--545},
  doi = {10.1038/313541a0}
}

@Article{ho24,
  author = 	 {Ho, W.-G. and Zhang, P. and Haule, K. and Jackson, J. M. and  Dobrosavljevi\'c, V. and  Dobrosavljevi\'c, V. V.},
  title = 	 {Quantum critical phase of {FeO} spans conditions of {E}arth's lower mantle},
  journal = 	 {Nat. Commun.},
  year = 	 2024,
  volume =	 15,
  pages =	 {3461},
  doi = {10.1038/s41467-024-47489-w}
}

@Article{holme13,
  author = 	 {Holme, R. and de Viron, O.},
  title = 	 {Characterization and implications of intradecadal variations in length of day},
  journal = 	 {Nature},
  year = 	 2013,
  volume =	 499,
  pages =	 {202--204},
  doi = {10.1038/nature12282}
}

@Article{jackson00,
  author = 	 {Jackson, A. and Jonkers, A. R. T. and Walker, M. R.},
  title = 	 {Four centuries of geomagnetic secular variation from historical records},
  journal = PTRSLA,
  year = 	 2000,
  volume =	 358,
  pages =	 {957--990},
  doi = {10.1098/rsta.2000.0569}
}

@Article{jault08,
  author = 	 {Jault, D.},
  title = 	 {Axial invariance of rapidly varying diffusionless motions in the {E}arth's core interior},
  journal = 	 PEPI,
  year = 	 2008,
  volume = 	 166,
  pages = 	 {67--76},
  doi = {10.1016/j.pepi.2007.11.001}
}

@Article{jault15,
  author = 	 {Jault, D.},
  title = 	 {Illuminating the electrical conductivity of the lowermost mantle from below},
  journal =  GJI,
  year = 	 2015,
  volume = 202,
  pages = 	 {482--496},
  doi = {10.1093/gji/ggv152}
}

@Article{lecomte23,
  author = 	 {Lecomte, H. and Rosat, S. and Mandea, M. and Dumberry, M.},
  title = 	 {Gravitational constraints on the {E}arth's inner core differential rotation},
  journal = GRL,
  year = 	 2023,
  volume = 	 {50},
  pages = 	 {e2023GL104790},
  doi = {10.1029/2023GL104790}
}

@Article{luo22a,
  author = 	 {Luo, J. and Jackson, A.},
  title = 	 {Waves in the {E}arth's core. {I}. {M}ildly diffusive torsional oscillations},
  journal = 	 {Proc. R. Soc. A},
  year = 	 2022,
  volume =	 478,
  pages =	 {20210982},
  doi = {10.1098/rspa.2021.0982}
}

@Article{mcnamara19,
  author = 	 {McNamara, A. K.},
  title = 	 {A review of large low shear velocity provinces and ultra-low velocity zones},
  journal = {Tectonophysics},
  year = 	 2019,
  volume =	 760,
  pages =	 {199--220},
  doi = {10.1016/j.tecto.2018.04.015}
}

@Article{mound03,
  author = 	 {Mound, J. E. and Buffett, B. A.},
  title = 	 {Interannual oscillations in the length of day: implications for the structure of mantle and core},
  journal = 	 JGR,
  year = 	 2003,
  volume =	 {108(B7)},
  pages =	 {2334},
  doi = {10.1029/2002jb002054}
}

@Article{mound05b,
  author = 	 {Mound, J. E. and Buffett,  B. A.},
  title = 	 {Mechanisms of core-mantle angular momentum exchange and the observed spectral properties of torsional oscillations},
  journal = 	 JGR,
  year = 	 2005,
  volume =	 110,
  pages =	 {B08103},
  doi = {10.1029/2004jb003555}
}

@Article{mound06,
  author = 	 {Mound, J. E. and Buffett, B. A.},
  title = 	 {Detection of a gravitational oscillation in length-of-day},
  journal = 	 EPSL,
  year = 	 2006,
  volume =	 243,
  pages =	 {383--389},
  doi = {10.1016/j.epsl.2006.01.043}
}

@Article{mound07,
  author = 	 {Mound, J. E. and Buffett, B. A.},
  title = 	 {Viscosity of the {E}arth's fluid core and torsional oscillations},
  journal = 	 {J. Geophys. Res.},
  year = 	 2007,
  volume =	 112,
  pages =	 {B05402},
  doi = {10.1029/2006JB004426}
}

@Article{ohta12,
  author = 	 {Ohta, K. and Cohen, R. E. and Hirose,  K. and Haule, K. and Shimiziu, K. and Ohishi, Y.},
  title = 	 {Experimental and theoretical evidence for pressure-induced metallization
in {FeO} with rocksalt-type structure},
  journal = 	 PRL,
  year = 	 2012,
  volume =	 108,
  pages =	 {026403},
  doi = {10.1103/physrevlett.108.026403}
}

@Article{pfeffer23,
  author = 	 {Pfeffer, J. and Cazenave, A. and Rosat, S. and Moreira, L. and Mandea, M. and  Dehant, V. and Coupry, B.},
  title = 	 {A 6-year cycle in the {E}arth system},
  journal = 	 {Glob. Planet. Change},
  year = 	 2023,
  volume =	 229,
  pages =	 {104245},
  doi = {10.1016/j.gloplacha.2023.104245}
}

@Article{russell23,
  author = 	 {Russell, S. and Irving, J. and Jagt, L. and Cottaar, S.},
  title = 	 {Evidence for a kilometer-scale seismically slow layer atop the core-mantle boundary from normal modes},
  journal = 	 GRL,
  year = 	 2023,
  volume =	 50,
  pages =	 {e2023GL105684},
  doi = {10.1029/2023gl105684}
}

@Article{schaeffer16,
author =  {Schaeffer, N. and Jault, D.},
year =    2016,
title = {Electrical conductivity of the lowermost mantle explains absorption of core torsional waves at the equator},
journal = GRL,
volume = 43,
pages = {4922--4928},
  doi = {10.1002/2016gl068301}
}

@Article{shih21,
  author = 	 {Shih, S. A. and Chao, B. F.},
  title = 	 {Inner core and its libration under gravitational equilibrium: implications to lower mantle density anomaly},
  journal = 	 {J. Geophys. Res. Solid Earth},
  year = 	 2021,
  volume =	 126,
  pages =	 {e2020JB020541},
  doi = {10.1029/2020JB020541}
}

@Article{simmons07,
  author = 	 {Simmons, N. A. and Forte, A. M. and Grand, S. P.},
  title = 	 {Thermochemical structure and dynamics of the {A}frican superplume},
  journal = 	 GRL,
  year = 	 2007,
  volume =	 34,
  doi = {10.1029/2006gl028009}
}

@Article{taylor63,
  author = 	 {Taylor, J. B.},
  title = 	 {The magneto-hydrodynamics of a rotating fluid and the {E}arth's dynamo problem},
  journal = 	 PRSLA,
  year = 	 1963,
  volume =	 274,
  pages =	 {274-283},
  doi = {10.1098/rspa.1963.0130}
}

@Article{wang22,
  author = 	 {Wang, W. and Vidale, J. E.},
  title = 	 {Seismological observation of {E}arth's oscillating inner core},
  journal = 	 {Sci. Adv.},
  year = 	 2022,
  volume =	 8,
  pages =	 {eabm9916},
  doi = {10.1126/sciadv.abm9916}
}

@Article{wang24,
  author = 	 {Wang, W. and Vidale, J. E. and Pang, G. and Koper, K. D. and Wang, R.},
  title = 	 {Inner core backtracking by seismic waveform change reversals},
  journal = 	 {Nature},
  year = 	 2024,
  volume =	 631,
  pages =	 {340--343},
  doi = {10.1038/s41586-024-07536-4}
}

@Article{zatman97,
  author = 	 {Zatman, S. and Bloxham, J.},
  title = 	 {Torsional oscillations and the magnetic field within the {E}arth's core},
  journal = 	 {Nature},
  year = 	 1997,
  volume =	 388,
  pages =	 {760--763},
  doi = {10.1038/41987}
}

\end{document}